\PassOptionsToPackage{unicode}{hyperref}
\PassOptionsToPackage{hyphens}{url}
\PassOptionsToPackage{dvipsnames,svgnames,x11names}{xcolor}
\documentclass[
  10pt,
  letterpaper,
]{article}
\usepackage{xcolor}
\usepackage[margin=1in]{geometry}
\usepackage{amsmath,amssymb}
\usepackage{iftex}
\ifPDFTeX
  \usepackage[T1]{fontenc}
  \usepackage[utf8]{inputenc}
  \usepackage{textcomp} 
\else 
  \usepackage{unicode-math} 
  \defaultfontfeatures{Scale=MatchLowercase}
  \defaultfontfeatures[\rmfamily]{Ligatures=TeX,Scale=1}
\fi
\usepackage{lmodern}
\ifPDFTeX\else
\fi
\IfFileExists{upquote.sty}{\usepackage{upquote}}{}
\IfFileExists{microtype.sty}{
  \usepackage[]{microtype}
  \UseMicrotypeSet[protrusion]{basicmath} 
}{}
\makeatletter
\@ifundefined{KOMAClassName}{
  \IfFileExists{parskip.sty}{%
    \usepackage{parskip}
  }{
    \setlength{\parindent}{0pt}
    \setlength{\parskip}{6pt plus 2pt minus 1pt}}
}{
  \KOMAoptions{parskip=half}}
\makeatother
\usepackage{color}
\usepackage{fancyvrb}

\DefineVerbatimEnvironment{Highlighting}{Verbatim}{commandchars=\\\{\}}
\newenvironment{Shaded}{}{}

\newcommand{\DataTypeTok}[1]{\textcolor[rgb]{0.56,0.13,0.00}{#1}}
\newcommand{\DecValTok}[1]{\textcolor[rgb]{0.25,0.63,0.44}{#1}}

\newcommand{\FunctionTok}[1]{\textcolor[rgb]{0.02,0.16,0.49}{#1}}

\newcommand{\NormalTok}[1]{#1}

\newcommand{\OtherTok}[1]{\textcolor[rgb]{0.00,0.44,0.13}{#1}}

\newcommand{\StringTok}[1]{\textcolor[rgb]{0.25,0.44,0.63}{#1}}

\usepackage{longtable,booktabs,array}
\usepackage{calc} 
\usepackage{etoolbox}
\makeatletter
\patchcmd\longtable{\par}{\if@noskipsec\mbox{}\fi\par}{}{}
\makeatother
\IfFileExists{footnotehyper.sty}{\usepackage{footnotehyper}}{\usepackage{footnote}}
\makesavenoteenv{longtable}
\usepackage{graphicx}
\makeatletter
\newsavebox\pandoc@box
\newcommand*\pandocbounded[1]{
  \sbox\pandoc@box{#1}%
  \Gscale@div\@tempa{\textheight}{\dimexpr\ht\pandoc@box+\dp\pandoc@box\relax}%
  \Gscale@div\@tempb{\linewidth}{\wd\pandoc@box}%
  \ifdim\@tempb\p@<\@tempa\p@\let\@tempa\@tempb\fi
  \ifdim\@tempa\p@<\p@\scalebox{\@tempa}{\usebox\pandoc@box}%
  \else\usebox{\pandoc@box}%
  \fi%
}
\def\fps@figure{htbp}
\makeatother
\providecommand{\tightlist}{%
  \setlength{\itemsep}{0pt}\setlength{\parskip}{0pt}}
\usepackage{bookmark}
\IfFileExists{xurl.sty}{\usepackage{xurl}}{} 
\hypersetup{
  pdftitle={Resident KV Claims: A Conformance Contract for Future Reuse under Active KV Pressure},
  pdfauthor={Lukas Stepanek},
  colorlinks=true,
  linkcolor={blue},
  filecolor={Maroon},
  citecolor={Blue},
  urlcolor={blue},
  pdfcreator={LaTeX via pandoc}}

\title{Resident KV Claims: A Conformance Contract for Future Reuse under
Active KV Pressure}
\author{Lukas Stepanek\\\texttt{luki.step@proton.me}}
\date{May 2026}

\begin{document}
\maketitle

\subsection{Abstract}\label{abstract}

KV-cache reuse mechanisms increasingly expose priority, duration,
offload, routing hints, scheduler modes, and event streams. These
mechanisms help preserve reusable prefixes, but they do not by
themselves define a portable contract for accepted future-reuse state
when resident KV and active live KV cannot both fit. We introduce
\textbf{resident KV claims}, a conformance contract that binds
future-reuse intent to a materialization predicate, lifecycle state,
active/resident feasibility outcome, and claim-level telemetry.

In controlled vLLM allocator probes, a 60-block resident claim and a
70-block active prefill exceed an 80-block usable KV pool. Write
no-admit prevents the active request from becoming future reusable
state, but it still allows active allocation to evict residents from the
shared pool. A minimal vLLM prototype shows that hard protected resident
claims convert this failure mode into scheduler-visible active refusal
with direct blocking-claim attribution.

The result is not a production speedup or a new cache-replacement
algorithm. It is a runtime contract that turns unreported resident loss
into reconstructable active/resident arbitration. A companion
MicroRuntime and vLLM litmus suite distinguish ordinary eviction, soft
priority, write no-admit, accepted hard claims, materialization failure,
demotion, expiry, active refusal, and trace-level outcome
reconstruction.

\subsection{1. Introduction}\label{introduction}

KV-cache reuse has become a central serving optimization for
decoder-only language models. Prefix caching avoids recomputing
attention keys and values for repeated prompt prefixes. Paged KV
allocators reduce fragmentation. Radix trees and priority-aware eviction
policies improve the chance that valuable prefixes survive until reuse.
Recent production runtimes and orchestration layers also expose stronger
controls, including token-range retention priority, duration hints,
offload, no-evict scheduler modes, routing metadata, and reuse telemetry
{[}1,4,5,9{]}.

These mechanisms make future reuse actionable, but they leave open the
contract question this paper studies:

\textbf{What should a runtime do when future-reuse state that should
remain resident and active live KV needed by the current request cannot
both fit?}

This paper's thesis is that existing KV-retention mechanisms expose
policies, while resident KV claims define the missing conformance
contract for what a runtime must report or do when accepted future-reuse
state conflicts with active live KV.

\begin{figure}
\centering
\includegraphics[width=0.95\linewidth,height=\textheight,keepaspectratio,alt={Resident Claim Thesis}]{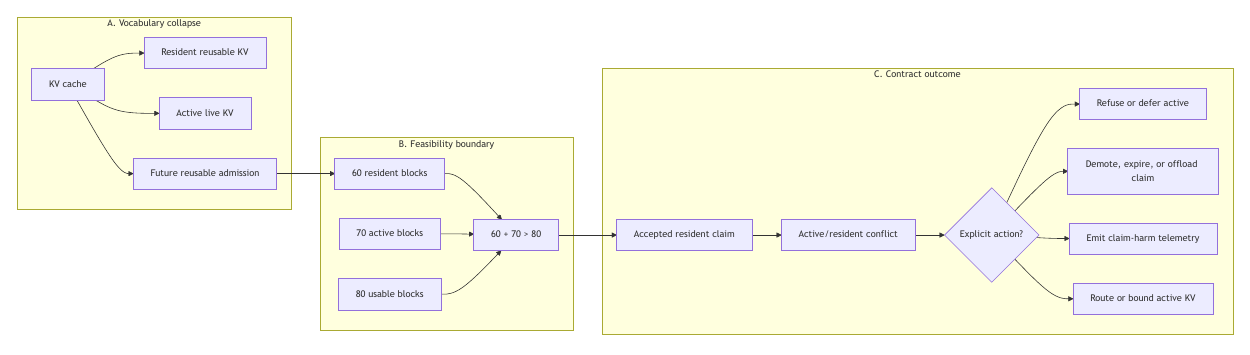}
\caption{Resident Claim Thesis}
\end{figure}

Figure 1 summarizes the contribution. Resident KV claims make
active/resident infeasibility observable rather than implicitly reducing
accepted future-reuse state to ordinary cache eviction.

The common vocabulary of ``KV cache'' obscures three different
resources:

\begin{enumerate}
\def\labelenumi{\arabic{enumi}.}
\tightlist
\item
  \textbf{Resident reusable KV:} KV already computed, not currently
  active, and valuable only if enough leading prefix state survives
  until a future request.
\item
  \textbf{Active live KV:} KV required to execute an in-flight request.
  Under full attention, active prefill chunks accumulate unless earlier
  chunks are freed, offloaded, or recomputed.
\item
  \textbf{Future reusable admission:} The decision to insert newly
  produced active KV into a reusable prefix cache after or during
  service.
\end{enumerate}

The distinction matters because a write no-admit policy acts on the
third resource. This paper uses \textbf{write no-admit} for a
per-request admission-control behavior: the active request may be
served, but the KV it produces is not admitted as future reusable cache
state. This can prevent a bulky request from becoming reusable in the
future. It does not automatically make the active request consume zero
KV while it is being served. If active allocation draws from the same
physical pool as resident reusable KV, no-admit can be too late to
protect residents.

Resident claims are not another eviction priority. A priority ranks
victims when eviction is allowed. A resident claim defines when eviction
is no longer ordinary cache replacement: after acceptance,
predicate-breaking loss must be preceded by demotion, expiry, offload,
refusal, or harm telemetry. This distinction matters precisely when
active live KV and protected resident KV cannot both fit.

This paper makes three contributions:

\begin{enumerate}
\def\labelenumi{\arabic{enumi}.}
\tightlist
\item
  It defines a resident-claim conformance contract with accepted,
  refused, demoted, expired, harmed, and materialized states; an
  active/resident feasibility check; explicit conflict actions; and
  claim-level telemetry.
\item
  It provides a runnable litmus suite showing that retained block count
  is not the right value unit, that write no-admit controls future
  reusable admission rather than active live allocation, and that
  post-demotion or post-expiry block loss is not claim harm.
\item
  It implements a minimal vLLM contract hook with resident-claim
  metadata, hard protected-resident victim exclusion, predicate-level
  lifecycle telemetry, and scheduler-visible active refusal with direct
  blocking-claim attribution, then bounds that result against runtime
  and workflow-serving prior art.
\end{enumerate}

The claim boundary is deliberate. This paper does not show a production
serving speedup, p95 or p99 latency improvement, a universal allocator,
a learned prediction model, or cross-runtime empirical superiority. TTFT
measurements are used only to show that prefix reuse has serving-visible
value, not that the arbiter improves end-to-end serving performance.
This is a mechanisms paper: it identifies a real runtime boundary,
falsifies simpler explanations, implements the minimal contract
behavior, and bounds the missing abstraction.

\subsection{2. Background And Prior-Art
Boundary}\label{background-and-prior-art-boundary}

\subsubsection{2.1 Prefix Caching and Paged
KV}\label{prefix-caching-and-paged-kv}

In decoder-only LLM serving, each generated token depends on attention
over earlier tokens. Serving systems therefore store per-layer key and
value tensors in KV caches. Prefix caching reuses KV for repeated prompt
prefixes, avoiding prefill recomputation when a future request shares a
prefix with an earlier one {[}2{]}.

PagedAttention and related block allocators divide KV memory into blocks
and manage them with page-like indirection {[}1{]}. This enables more
flexible allocation than one contiguous tensor per request. It also
creates a natural block-level cache replacement problem: which free,
reusable blocks should be evicted when capacity is needed?

\subsubsection{2.2 Retention Primitives Are Real Prior
Art}\label{retention-primitives-are-real-prior-art}

The key prior-art fact is that modern runtimes already expose
KV-retention fragments.

TensorRT-LLM is the closest runtime comparator. Its public material
describes priority-based KV eviction, token-range retention priority,
optional duration, KV cache events, prioritized LRU, primary and
secondary memory behavior, and offload support {[}4,5{]}. Its runtime
flags also include scheduler modes such as guaranteed no-evict and
max-utilization modes {[}6{]}.

SGLang exposes a radix-cache design, lock references, protected and
evictable accounting, eviction policy choices, and HiCache
offload/storage paths {[}7,8{]}. Dynamo's SGLang agentic-workload guide
also documents priority scheduling, priority-based radix eviction,
hierarchical-cache interaction, and experimental session control for KV
isolation {[}10{]}.

vLLM exposes prefix caching, block-pool management, request scheduling,
and an emerging public discussion around context-aware KV retention APIs
{[}3{]}. The vLLM 0.19.1 prototype uses that runtime as the live
counterexample, not as evidence that vLLM lacks future retention
primitives.

The vLLM retention RFC is especially close to this work because it
treats a retention API as the product, proposes token-range priority and
duration, and separates request scheduling priority from block eviction
priority. The distinction here is that resident claims define
accepted-claim conformance: materialization predicates, active/resident
infeasibility outcomes, and claim-harm/refusal telemetry.

Dynamo, Continuum, KVFlow, Pie, Marconi, and vLLM's Mooncake integration
sharpen the same boundary from adjacent directions. Dynamo describes
agent hints, scheduling and routing metadata, backend-specific priority
scheduling and priority-based eviction, speculative prefill, session
control, and KV-event-driven orchestration {[}9,10{]}. Continuum treats
retained KV during tool-call gaps as a scheduling and GPU-memory
tradeoff {[}11{]}. Continuum is therefore the closest agentic retention
policy comparator, but it does not subsume the contract studied here: a
retention policy can decide what to keep, while a resident claim defines
what an accepted promise means under active-live infeasibility. KVFlow
uses workflow structure to guide future reuse and prefetch {[}12{]}. Pie
exposes programmable serving handlers that can implement
application-specific KV strategies {[}13{]}. Marconi frames admission
and eviction around reuse likelihood and compute-saving value per
footprint {[}14{]}. The Mooncake integration pushes vLLM toward
distributed KV storage for agentic workloads {[}15{]}.

The baseline is not a runtime with no retention machinery. Existing
systems plainly expose priority, TTL, offload, no-evict, routing,
programmability, and prefix-cache mechanisms. The distinction is
semantic: a priority, duration, lock reference, offload tier, or routing
hint becomes a ResidentClaim implementation only if it preserves the
accepted-claim lifecycle, materialization predicate, active/resident
conflict outcome, and claim-level telemetry required by the contract.

\begin{longtable}[]{@{}
  >{\raggedright\arraybackslash}p{(\linewidth - 4\tabcolsep) * \real{0.3333}}
  >{\raggedright\arraybackslash}p{(\linewidth - 4\tabcolsep) * \real{0.3333}}
  >{\raggedright\arraybackslash}p{(\linewidth - 4\tabcolsep) * \real{0.3333}}@{}}
\caption{Prior-art boundary for resident KV claims.}\tabularnewline
\toprule\noalign{}
\begin{minipage}[b]{\linewidth}\raggedright
System or mechanism
\end{minipage} & \begin{minipage}[b]{\linewidth}\raggedright
What it provides
\end{minipage} & \begin{minipage}[b]{\linewidth}\raggedright
Why it is not the full ResidentClaim contract
\end{minipage} \\
\midrule\noalign{}
\endfirsthead
\toprule\noalign{}
\begin{minipage}[b]{\linewidth}\raggedright
System or mechanism
\end{minipage} & \begin{minipage}[b]{\linewidth}\raggedright
What it provides
\end{minipage} & \begin{minipage}[b]{\linewidth}\raggedright
Why it is not the full ResidentClaim contract
\end{minipage} \\
\midrule\noalign{}
\endhead
\bottomrule\noalign{}
\endlastfoot
TensorRT-LLM priority / duration & Token-range priority/duration,
priority-aware eviction, KV events & Soft-retention substrate; does not
define accepted-claim lifecycle, predicate harm, active refusal, or
blocking-claim attribution. \\
TensorRT-LLM guaranteed no-evict & No-evict protection for an active
running request & Active-request scoped; not future-resident reusable KV
under later pressure, and not claim materialization/refusal/harm
semantics. \\
SGLang / HiCache & Radix cache, lock refs, protected/evictable
accounting, tiers, prefetch/write-back & Storage and prefix-cache
substrate; public semantics do not define accepted claims, predicate
restoration failure, or claim-scoped conflicts. \\
Dynamo / orchestration layers & Agent hints, routing metadata, KV-aware
scheduling/orchestration & Can route around conflicts, but metadata is
not backend-local proof of accepted, materialized, preserved, demoted,
harmed, or blocking claims. \\
vLLM retention RFC / retention APIs & API pressure toward priority-style
retention controls & Shows the problem is live; this paper studies
future-reuse admission versus active live allocation and explicit
accepted-claim conflicts. \\
This vLLM prototype & Metadata, materialization, hard exclusion,
demotion/expiry, active refusal telemetry & Patch-level reference for
the contract subset; not an upstream API, production policy, or
performance claim. \\
\end{longtable}

This boundary is why the contribution is a conformance contract rather
than a new eviction heuristic. Existing mechanisms are plausible
lowering targets. This paper defines the obligations they would need to
satisfy.

\subsubsection{2.3 Why ``Cache Eviction'' Is Too
Small}\label{why-cache-eviction-is-too-small}

Classical cache replacement asks which cached object to evict when
another cached object should be admitted. Predictive KV residency
includes that question, but it also includes a different one: should an
active request be served now if doing so consumes memory that was
promised to resident future-reuse state?

If the runtime has 80 usable KV blocks, 60 protected resident blocks,
and a new active full-attention prefill needing 70 live blocks, then no
replacement ranking can make all desired state fit:

\begin{Shaded}
\begin{Highlighting}[]
\NormalTok{protected resident KV = 60 blocks}
\NormalTok{active live KV        = 70 blocks}
\NormalTok{usable KV pool        = 80 blocks}

\NormalTok{60 + 70 = 130 \textgreater{} 80}
\end{Highlighting}
\end{Shaded}

Something must give. A runtime can evict or relax residents, defer or
preempt active work, offload state, bound active live KV, split or
recompute, route to another worker, reject reusable admission, or add
capacity. But it cannot preserve all resident KV and serve the active
request in the same 80-block pool without some such action.

\subsection{3. Contract}\label{contract}

\subsubsection{3.1 Definitions}\label{definitions}

\textbf{Resident reusable KV} is previously computed KV that is not
needed by an active request but may be reused by a future request. It
has value only through a future materialization surface, such as a
leading-prefix hit.

\textbf{Active live KV} is KV required to execute an in-flight request.
In full attention, active live KV grows across chunks because later
chunks attend to earlier chunks. Chunked scheduling reduces a compute
burst; it does not by itself cap live KV.

\textbf{Future reusable admission} is the decision to retain newly
produced active KV as future reusable state.

\textbf{Useful resident value} is policy-defined. In the evidence below,
a prefix receives value only when its leading contiguous survival
crosses a threshold. This is intentionally narrower than ``some cached
tokens survived.'' Below-threshold retained tokens may save some compute
in another policy, but they fail the intended resident claim.

\textbf{Resident claim} is an application-visible future-reuse object
submitted to the runtime. In this paper, a claim names the fields
required for conformance: a stable claim id, owner scope,
cache-equivalence identity, prefix object, materialization predicate,
footprint estimate, protection mode, and optional duration. Value,
confidence, budgets, deadlines, and overbooking policy are optional
policy inputs, not required parts of the conformance contract.

\textbf{Accepted resident claim} means the runtime has taken
responsibility for preserving the claim's materialization predicate
unless it emits an explicit refusal, demotion, offload, expiry, or harm
event. This is stronger than ``high-priority cache entry'' and weaker
than an unconditional guarantee that ignores physical capacity.

A runtime is allowed to reject a resident claim. The conformance
obligation becomes binding only after the runtime accepts the claim.

\textbf{Claim harm} means an accepted claim lost enough state to fail
its materialization predicate before the runtime emitted an acceptable
demotion, expiry, or refusal. Ordinary cache eviction is not claim harm
unless the runtime had accepted such a claim.

\textbf{Active/resident arbiter} is the runtime component or contract
that decides what happens when active live KV and resident reusable KV
conflict.

The minimal conformance schema is deliberately split so that inputs,
decisions, lifecycle state, and telemetry cannot blur together:

\begin{longtable}[]{@{}
  >{\raggedright\arraybackslash}p{(\linewidth - 4\tabcolsep) * \real{0.3333}}
  >{\raggedright\arraybackslash}p{(\linewidth - 4\tabcolsep) * \real{0.3333}}
  >{\raggedright\arraybackslash}p{(\linewidth - 4\tabcolsep) * \real{0.3333}}@{}}
\caption{Minimal resident-claim conformance schema.}\tabularnewline
\toprule\noalign{}
\begin{minipage}[b]{\linewidth}\raggedright
Surface
\end{minipage} & \begin{minipage}[b]{\linewidth}\raggedright
Required fields
\end{minipage} & \begin{minipage}[b]{\linewidth}\raggedright
Outside the conformance contract
\end{minipage} \\
\midrule\noalign{}
\endfirsthead
\toprule\noalign{}
\begin{minipage}[b]{\linewidth}\raggedright
Surface
\end{minipage} & \begin{minipage}[b]{\linewidth}\raggedright
Required fields
\end{minipage} & \begin{minipage}[b]{\linewidth}\raggedright
Outside the conformance contract
\end{minipage} \\
\midrule\noalign{}
\endhead
\bottomrule\noalign{}
\endlastfoot
\texttt{ResidentClaimInput} & \texttt{claim\_id}, \texttt{owner\_scope},
\texttt{cache\_identity}, \texttt{object\_id},
\texttt{materialization\_predicate}, \texttt{footprint\_blocks},
\texttt{protection\_mode}, optional \texttt{duration\_steps} &
\texttt{value}, \texttt{confidence}, \texttt{budget}, \texttt{deadline},
and policy scores \\
\texttt{CacheIdentity} & cache-key domain, model id, tokenizer or
token-hash domain, namespace or salt, block size, optional
adapter/prompt-embedding identity and KV format & workload-specific
prediction metadata \\
\texttt{ResidentClaimDecision} & accepted, rejected, or conditionally
accepted; decision step; reason & claim outcome events \\
\texttt{ResidentClaimState} & submitted, accepted, materialized,
demoted, expired, refused, harmed & admission-policy inputs \\
\texttt{ClaimEvent} & event type, claim id, step, optional request id,
predicate/capacity fields & raw block events without claim context \\
\end{longtable}

\texttt{protection\_mode} is normative, not a generic ``hardness''
label:

\begin{longtable}[]{@{}
  >{\raggedright\arraybackslash}p{(\linewidth - 2\tabcolsep) * \real{0.5000}}
  >{\raggedright\arraybackslash}p{(\linewidth - 2\tabcolsep) * \real{0.5000}}@{}}
\caption{Resident-claim protection modes.}\tabularnewline
\toprule\noalign{}
\begin{minipage}[b]{\linewidth}\raggedright
Mode
\end{minipage} & \begin{minipage}[b]{\linewidth}\raggedright
Meaning
\end{minipage} \\
\midrule\noalign{}
\endfirsthead
\toprule\noalign{}
\begin{minipage}[b]{\linewidth}\raggedright
Mode
\end{minipage} & \begin{minipage}[b]{\linewidth}\raggedright
Meaning
\end{minipage} \\
\midrule\noalign{}
\endhead
\bottomrule\noalign{}
\endlastfoot
\texttt{soft\_priority} & May influence eviction order; not a hard
claim. \\
\texttt{hard\_protected} & May not be broken before explicit active
refusal/defer, demotion, expiry, offload, or harm event. \\
\texttt{demotable} & May be downgraded before loss, but the demotion
event must precede predicate-breaking eviction. \\
\texttt{offloadable} & May leave GPU memory only if the materialization
predicate remains restorable. \\
\texttt{expiring} & Runtime responsibility ends at expiry. \\
\texttt{best\_effort} & Telemetry only; no preservation obligation. \\
\end{longtable}

\subsubsection{3.2 Claim Harm Semantics}\label{claim-harm-semantics}

Claim harm is a lifecycle violation, not a synonym for cache eviction.
The same block loss can be harmless, expected, or contract-breaking
depending on whether the runtime had accepted a claim and whether it
released the claim first.

\begin{longtable}[]{@{}
  >{\raggedright\arraybackslash}p{(\linewidth - 4\tabcolsep) * \real{0.3333}}
  >{\raggedright\arraybackslash}p{(\linewidth - 4\tabcolsep) * \real{0.3333}}
  >{\raggedright\arraybackslash}p{(\linewidth - 4\tabcolsep) * \real{0.3333}}@{}}
\caption{Claim-harm semantics by event sequence.}\tabularnewline
\toprule\noalign{}
\begin{minipage}[b]{\linewidth}\raggedright
Event sequence
\end{minipage} & \begin{minipage}[b]{\linewidth}\raggedright
Interpretation
\end{minipage} & \begin{minipage}[b]{\linewidth}\raggedright
Required telemetry
\end{minipage} \\
\midrule\noalign{}
\endfirsthead
\toprule\noalign{}
\begin{minipage}[b]{\linewidth}\raggedright
Event sequence
\end{minipage} & \begin{minipage}[b]{\linewidth}\raggedright
Interpretation
\end{minipage} & \begin{minipage}[b]{\linewidth}\raggedright
Required telemetry
\end{minipage} \\
\midrule\noalign{}
\endhead
\bottomrule\noalign{}
\endlastfoot
Ordinary cached block evicted without accepted claim & Normal cache
replacement & Block removal may be reported, but not
\texttt{claim\_harmed}. \\
Accepted claim is demoted, then blocks are lost & Policy explicitly
relaxed responsibility before loss & \texttt{claim\_demoted} before
post-release block loss. \\
Accepted claim expires, then blocks are lost & Responsibility ended by
duration or expiry rule & \texttt{claim\_expired} before post-release
block loss. \\
Accepted claim loses its materialization predicate without prior release
& True claim harm & \texttt{claim\_harmed} with predicate and capacity
context. \\
Accepted hard claim blocks active admission & Resident claim preserved;
active side takes the action & \texttt{active\_refused} or
\texttt{active\_deferred} with \texttt{blocking\_claim\_ids}. \\
\end{longtable}

This distinction is the observability core of the contract. Raw KV
events that say a block was stored, updated, or removed are useful, but
they do not by themselves tell an external observer whether the runtime
violated an accepted future-reuse responsibility.

\subsubsection{3.3 Claim Lifecycle}\label{claim-lifecycle}

The proposed unit is a claim over a materializable future computation
state, not an individual cache block. A minimal lifecycle is:

\begin{figure}
\centering
\includegraphics[width=0.95\linewidth,height=\textheight,keepaspectratio,alt={Resident claim lifecycle}]{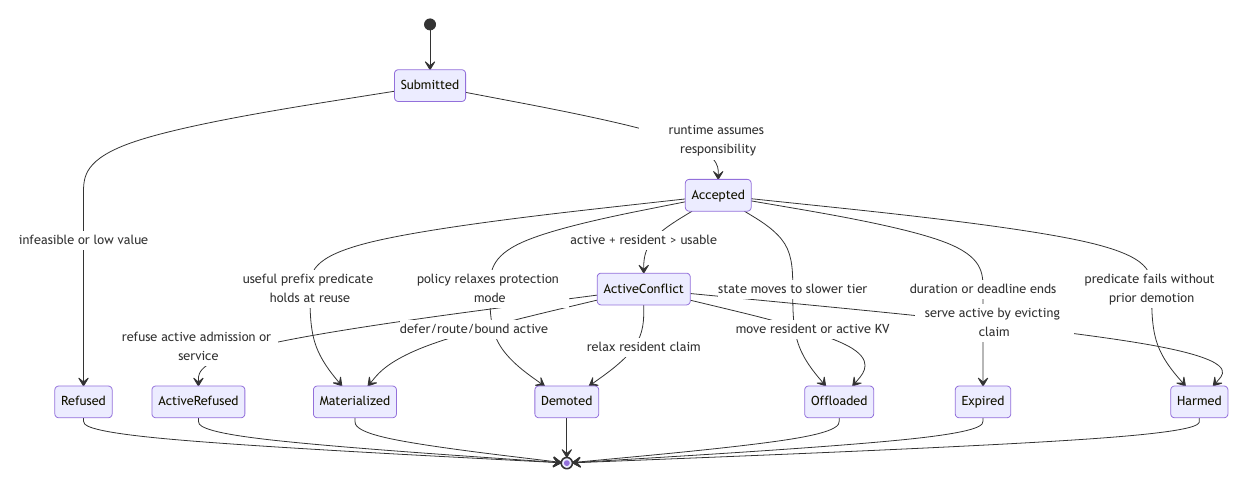}
\caption{Resident claim lifecycle}
\end{figure}

\subsubsection{3.4 Feasibility Boundary}\label{feasibility-boundary}

The boundary is:

\begin{Shaded}
\begin{Highlighting}[]
\NormalTok{protected\_resident\_kv + active\_live\_kv \textless{}= usable\_kv}
\end{Highlighting}
\end{Shaded}

If the inequality holds, the runtime may serve active work and preserve
residents in the same pool, subject to ordinary eviction and scheduling
details.

If the inequality does not hold, a future-reuse hint cannot be treated
as an unconditional command. The runtime must choose an explicit action
and should report that action at the claim level.

\begin{figure}
\centering
\includegraphics[width=0.95\linewidth,height=\textheight,keepaspectratio,alt={Active resident KV arbitration}]{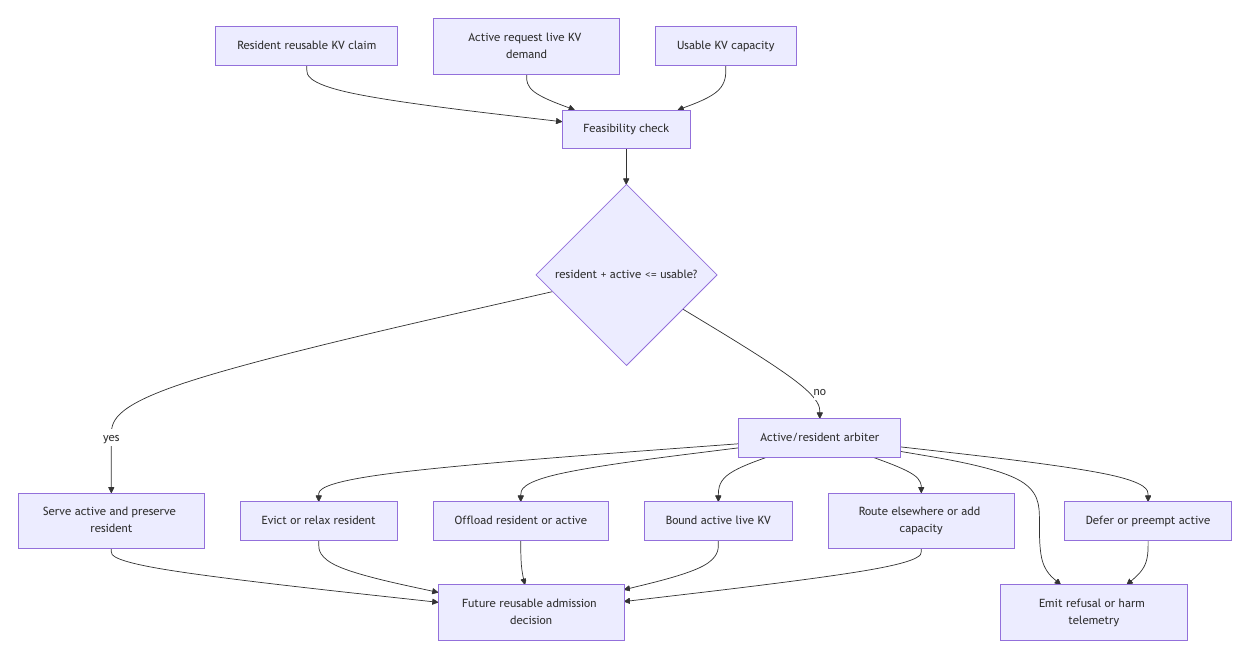}
\caption{Active resident KV arbitration}
\end{figure}

\subsubsection{3.5 Mechanism Table}\label{mechanism-table}

The table organizes the main contract actions for the 60/70/80 scenario.
It is derived from the MicroRuntime arbiter scenario with 60 resident
blocks, 70 active-live blocks, and 80 usable blocks. Costs are schematic
mechanism costs, not production latency measurements. Rows marked ``not
implemented in prototype'' are included to show valid
conflict-resolution actions under the contract; the evaluated prototype
implements only the rows marked implemented, observed, modeled, or sweep
baseline.

\begin{longtable}[]{@{}
  >{\raggedright\arraybackslash}p{(\linewidth - 4\tabcolsep) * \real{0.3333}}
  >{\raggedright\arraybackslash}p{(\linewidth - 4\tabcolsep) * \real{0.3333}}
  >{\raggedright\arraybackslash}p{(\linewidth - 4\tabcolsep) * \real{0.3333}}@{}}
\caption{Contract outcomes for active/resident conflict
mechanisms.}\tabularnewline
\toprule\noalign{}
\begin{minipage}[b]{\linewidth}\raggedright
Mechanism
\end{minipage} & \begin{minipage}[b]{\linewidth}\raggedright
Contract outcome
\end{minipage} & \begin{minipage}[b]{\linewidth}\raggedright
Prototype or evidence role
\end{minipage} \\
\midrule\noalign{}
\endfirsthead
\toprule\noalign{}
\begin{minipage}[b]{\linewidth}\raggedright
Mechanism
\end{minipage} & \begin{minipage}[b]{\linewidth}\raggedright
Contract outcome
\end{minipage} & \begin{minipage}[b]{\linewidth}\raggedright
Prototype or evidence role
\end{minipage} \\
\midrule\noalign{}
\endhead
\bottomrule\noalign{}
\endlastfoot
Native eviction & Active is served and reusable, but resident value is
lost. & Observed baseline; future-reuse hints become ordinary cached
state. \\
Write no-admit only & Active is served but not reusable; resident value
is still lost. & Implemented negative control showing that write
admission and active-live allocation are separate. \\
Resident victim exclusion & Resident survives; active is refused or
deferred when headroom is insufficient. & Implemented hard-protection
row. \\
Active deferral & Resident survives; active service is delayed and not
reusable until admitted. & Implemented scheduler-path refuse/defer
signal. \\
Resident reserve & Resident survives; active work that cannot fit is
refused. & Modeled reserve action, not a separate vLLM API. \\
Offload resident & Resident survives and active may be served, subject
to restore cost. & Valid contract action, not implemented in
prototype. \\
Offload active & Resident survives while active service pays bandwidth
or latency. & Valid contract action, not implemented in prototype. \\
Bound active-live KV & Resident and active may coexist if old chunks are
freed, offloaded, or recomputed. & Valid contract action; ordinary chunk
scheduling alone is insufficient. \\
Recompute or split & Memory pressure is traded for extra compute and
scheduling complexity. & Valid contract action, not implemented in
prototype. \\
Route elsewhere & Active is served on a worker with headroom; resident
survives locally. & Valid orchestration action requiring routing
visibility and spare capacity. \\
Relax resident claim & Active is served by demoting, expiring, or
partially evicting resident state. & Implemented through demotion and
expiry lifecycle tests. \\
Larger capacity & Resident and active coexist once the inequality is
satisfied. & Sweep baseline; capacity is not itself an allocation
contract. \\
Oracle upper bound & Best explicit action is chosen under known costs. &
Analysis row only, not an implementable policy without cost and
confidence inputs. \\
\end{longtable}

The table makes the thesis concrete: the proposed mechanism is not
``better LRU.'' It is the obligation to surface the conflict and choose
among these actions instead of collapsing accepted resident claims into
ordinary cache victims without claim-level accounting.

\subsection{4. Methods}\label{methods}

\subsubsection{4.1 Live vLLM Evidence}\label{live-vllm-evidence}

The vLLM evidence comes from controlled traces over three surfaces:
direct carrier traces for useful-prefix materialization, allocator-level
pressure probes for the 60 resident / 70 active / 80 usable boundary,
and \texttt{vllm.LLM.generate} pressure traces for scheduler-path
propagation. The strongest rows use a direct ledger rather than relying
only on native cache telemetry. The ledger checks that cached tokens
equal the first missing block times the 16-token block size, survived
positions form leading ranges, evicted blocks are not counted as
survived, block IDs map one-to-one to prefix positions, and
salted/no-reuse controls show zero useful leading-prefix survival.

\subsubsection{4.2 MicroRuntime Evidence}\label{microruntime-evidence}

The companion MicroRuntime is an executable contract model, not a claim
about production inference-engine behavior. It separates resident
reusable KV, active live KV, and future reusable admission so that the
claim schema, harm semantics, active-live accumulation, materialization
predicates, and arbiter action table can be tested independently of
vLLM-specific implementation details. The companion MicroRuntime
provides executable tests for the contract semantics; Appendix A lists
the relevant commands and artifacts.

\subsubsection{4.3 vLLM Prototype
Evidence}\label{vllm-prototype-evidence}

The vLLM prototype is a patch-level contract hook against vLLM base
commit \texttt{b1388b1}, not an upstream API. The published patch is
intentionally narrow and prototype-grade. It adds env-gated JSONL
telemetry, resident claim metadata, write no-admit for selected request
ids, protected-resident victim exclusion, claim relaxation/expiry/harm
events, scheduler-visible active refusal, and an engine-core path for
terminal scheduler outputs.

The reproducibility package consists of two commit-pinned public
repositories: the MicroRuntime executable model and the vLLM arbiter
artifact, including the runtime patch, generated traces, conformance
results, and evidence scripts. Detailed artifact paths are listed in
Appendix A.

\subsubsection{4.4 Evidence Summary}\label{evidence-summary}

The evidence is organized around a small number of deliberately narrow
points:

\begin{itemize}
\item
  \textbf{ResidentClaim conformance suite.} The vLLM arbiter conformance
  \texttt{results.json} reports seven passing trace/materialization
  checks, plus one executable capability-classification check that marks
  soft priority as an unsound lowering for hard protected claims. This
  gives the paper a conformance artifact rather than only prose
  definitions.
\item
  \textbf{Retained fragments below useful threshold.} The MicroRuntime
  materialization harness and controlled vLLM carrier trace show that
  naive fair share retained \texttt{480\ /\ 320\ /\ 304} cached tokens
  but produced thresholded value \texttt{0}; complete-prefix and
  value-density rows retained \texttt{640\ /\ 640\ /\ 0} and produced
  value \texttt{18}. Useful resident KV depends on materialization
  shape, not raw retained-block count.
\item
  \textbf{Active/resident feasibility boundary.} The native BlockPool,
  hard-claim, and conformance trace for accepted hard-claim
  infeasibility show that 60 resident blocks plus 70 active blocks
  exceed an 80-block usable pool. Ordinary allocation evicts residents;
  hard protection makes active allocation fail. The runtime therefore
  needs an explicit action when protected resident KV and active live KV
  cannot both fit.
\item
  \textbf{Write no-admit boundary.} The no-admit trace and MicroRuntime
  active-prefill tests show that bulky repeat reuse fell from
  \texttt{1120} cached tokens to \texttt{0} in the active-prefill model,
  while the vLLM no-admit trace still served active allocation through
  resident victims. Future reusable admission is separate from active
  live KV allocation.
\item
  \textbf{Chunking boundary.} The MicroRuntime \texttt{active\_live.py}
  harness and arbiter tests show that chunked schedule
  \texttt{20\ /\ 20\ /\ 20\ /\ 10} accumulates to 70 live active blocks
  under full attention. Chunked prefill does not bound live KV unless
  old chunks are freed, offloaded, or recomputed.
\item
  \textbf{Minimal vLLM arbiter prototype.} The capacity-sweep artifact
  shows that below 130 usable blocks, native and write no-admit serve
  active while losing resident materialization; hard resident exclusion
  preserves residents and refuses active with
  \texttt{blocking\_claim\_ids}. At and above 130 blocks, hard resident
  exclusion serves active while preserving residents. This changes
  allocator behavior in the predicted direction and attributes refusal
  to accepted resident claims.
\item
  \textbf{Live scheduler pressure path.} The live-scheduler-pressure
  summary shows a protected resident claim accepted with 40 resident
  blocks materialized; an active request requiring 46 blocks is deferred
  and then refused with blocking claim \texttt{claim:live-resident},
  combined resident-plus-active footprint of 86 blocks, a 19-block
  capacity shortfall, and protected-resident capacity refusal. This
  explicit active-side action exists in a real
  \texttt{vllm.LLM.generate} path, not only a direct BlockPool probe.
\item
  \textbf{Prefix reuse TTFT motivation.} The live-scheduler traces show
  repeated prompts seeing cached-token hits and lower TTFT after the
  first request. Preserved prefix reuse is serving-visible, but this is
  not evidence of an end-to-end arbiter performance gain.
\item
  \textbf{Prior-art boundary.} Public vLLM, SGLang, TensorRT-LLM,
  Dynamo, Continuum, KVFlow, Pie, Marconi, and Mooncake materials expose
  substantial KV-retention primitives. The contribution is the
  active/resident claim contract that specifies accepted-claim
  lifecycle, materialization, conflict outcomes, and claim-level
  telemetry.
\end{itemize}

\subsection{5. Question-Driven Results}\label{question-driven-results}

The evaluation is organized around four questions. Each question
corresponds to one failure mode that a resident-claim contract must
distinguish from ordinary cache behavior:

\begin{enumerate}
\def\labelenumi{\arabic{enumi}.}
\tightlist
\item
  Can retained block count diverge from useful materialization?
\item
  Does write no-admit protect resident KV under active pressure?
\item
  Does the active/resident feasibility boundary force explicit
  arbitration?
\item
  Can an observer reconstruct the accepted claim, conflict, and outcome
  from telemetry?
\end{enumerate}

\subsubsection{5.1 Q1: Is Useful Resident Value Equivalent To Retained
Block
Count?}\label{q1-is-useful-resident-value-equivalent-to-retained-block-count}

No.~The controlled carrier trace, which isolates materialization shape
from active-pressure effects, shows that raw retained-block count can
disagree with useful resident value.

\begin{longtable}[]{@{}
  >{\raggedright\arraybackslash}p{(\linewidth - 10\tabcolsep) * \real{0.1364}}
  >{\raggedleft\arraybackslash}p{(\linewidth - 10\tabcolsep) * \real{0.1818}}
  >{\raggedleft\arraybackslash}p{(\linewidth - 10\tabcolsep) * \real{0.1818}}
  >{\raggedleft\arraybackslash}p{(\linewidth - 10\tabcolsep) * \real{0.1818}}
  >{\raggedright\arraybackslash}p{(\linewidth - 10\tabcolsep) * \real{0.1364}}
  >{\raggedleft\arraybackslash}p{(\linewidth - 10\tabcolsep) * \real{0.1818}}@{}}
\caption{Retained tokens versus thresholded materialization
value.}\tabularnewline
\toprule\noalign{}
\begin{minipage}[b]{\linewidth}\raggedright
Policy
\end{minipage} & \begin{minipage}[b]{\linewidth}\raggedleft
A cached tokens
\end{minipage} & \begin{minipage}[b]{\linewidth}\raggedleft
B cached tokens
\end{minipage} & \begin{minipage}[b]{\linewidth}\raggedleft
C cached tokens
\end{minipage} & \begin{minipage}[b]{\linewidth}\raggedright
First missing blocks
\end{minipage} & \begin{minipage}[b]{\linewidth}\raggedleft
Thresholded value
\end{minipage} \\
\midrule\noalign{}
\endfirsthead
\toprule\noalign{}
\begin{minipage}[b]{\linewidth}\raggedright
Policy
\end{minipage} & \begin{minipage}[b]{\linewidth}\raggedleft
A cached tokens
\end{minipage} & \begin{minipage}[b]{\linewidth}\raggedleft
B cached tokens
\end{minipage} & \begin{minipage}[b]{\linewidth}\raggedleft
C cached tokens
\end{minipage} & \begin{minipage}[b]{\linewidth}\raggedright
First missing blocks
\end{minipage} & \begin{minipage}[b]{\linewidth}\raggedleft
Thresholded value
\end{minipage} \\
\midrule\noalign{}
\endhead
\bottomrule\noalign{}
\endlastfoot
Native & 0 & 0 & 0 & \texttt{0\ /\ 0\ /\ 0} & 0 \\
Naive fair share & 480 & 320 & 304 & \texttt{30\ /\ 20\ /\ 19} & 0 \\
Complete-prefix fair share & 640 & 640 & 0 & \texttt{40\ /\ 40\ /\ 0} &
18 \\
Value density & 640 & 640 & 0 & \texttt{40\ /\ 40\ /\ 0} & 18 \\
Salted/no-reuse & 0 & 0 & 0 & \texttt{0\ /\ 0\ /\ 0} & 0 \\
\end{longtable}

The important row is naive fair share. It retained 1,104 cached tokens
across A, B, and C, but failed the policy-defined useful threshold for
all spans. Complete-prefix fair share and value density retained fewer
total blocks than an infeasible ``save everything'' baseline, but
preserved the leading prefixes that carried value.

This result supports a narrow claim: useful predictive residency is not
equivalent to retaining arbitrary blocks. It does not support a
universal claim that below-threshold cached tokens save literally zero
compute under every possible policy.

\subsubsection{5.2 Q2: Does Write No-Admit Protect Resident
KV?}\label{q2-does-write-no-admit-protect-resident-kv}

No.~The active-prefill no-admit run separates future reusable admission
from active live allocation.

\begin{longtable}[]{@{}
  >{\raggedright\arraybackslash}p{(\linewidth - 10\tabcolsep) * \real{0.1304}}
  >{\raggedleft\arraybackslash}p{(\linewidth - 10\tabcolsep) * \real{0.1739}}
  >{\raggedleft\arraybackslash}p{(\linewidth - 10\tabcolsep) * \real{0.1739}}
  >{\raggedleft\arraybackslash}p{(\linewidth - 10\tabcolsep) * \real{0.1739}}
  >{\raggedleft\arraybackslash}p{(\linewidth - 10\tabcolsep) * \real{0.1739}}
  >{\raggedleft\arraybackslash}p{(\linewidth - 10\tabcolsep) * \real{0.1739}}@{}}
\caption{Effect of write no-admit on active reuse and resident
survival.}\tabularnewline
\toprule\noalign{}
\begin{minipage}[b]{\linewidth}\raggedright
Policy
\end{minipage} & \begin{minipage}[b]{\linewidth}\raggedleft
Bulky active served?
\end{minipage} & \begin{minipage}[b]{\linewidth}\raggedleft
Immediate bulky repeat reusable?
\end{minipage} & \begin{minipage}[b]{\linewidth}\raggedleft
Resident \texttt{small\_hot}
\end{minipage} & \begin{minipage}[b]{\linewidth}\raggedleft
Resident \texttt{small\_warm}
\end{minipage} & \begin{minipage}[b]{\linewidth}\raggedleft
Resident thresholded value
\end{minipage} \\
\midrule\noalign{}
\endfirsthead
\toprule\noalign{}
\begin{minipage}[b]{\linewidth}\raggedright
Policy
\end{minipage} & \begin{minipage}[b]{\linewidth}\raggedleft
Bulky active served?
\end{minipage} & \begin{minipage}[b]{\linewidth}\raggedleft
Immediate bulky repeat reusable?
\end{minipage} & \begin{minipage}[b]{\linewidth}\raggedleft
Resident \texttt{small\_hot}
\end{minipage} & \begin{minipage}[b]{\linewidth}\raggedleft
Resident \texttt{small\_warm}
\end{minipage} & \begin{minipage}[b]{\linewidth}\raggedleft
Resident thresholded value
\end{minipage} \\
\midrule\noalign{}
\endhead
\bottomrule\noalign{}
\endlastfoot
Cache-all active prefill & Yes & 1120 tokens & 0 & 0 & 0 \\
Served-but-not-reusable prefill & Yes & 0 tokens & 0 & 0 & 0 \\
Density-gated/write no-admit & Yes & 0 tokens & 0 & 0 & 0 \\
Salted/no-reuse & Yes & 0 tokens & 0 & 0 & 0 \\
\end{longtable}

No-admit succeeds at what it actually controls: the bulky request is not
reusable on immediate repeat. But no-admit fails as resident protection:
both compact resident prefixes still die. The mechanism is that active
allocation happens before, or independently from, reusable admission. A
shared physical pool still has to hold active KV while the request is
served.

This explains why a MicroRuntime no-cache prefill row should not be read
as an ordinary vLLM write-admission hook. In the model, no-cache prefill
is valid only under a stronger mechanism: separate active scratch space,
disposable active KV, offload, recomputation, or another action that
actually removes active live pressure from the resident pool.

Ordinary chunking is the same kind of trap. It schedules compute in
smaller pieces, but under full attention it does not bound the live KV
footprint unless older chunks are freed, offloaded, or recomputed.

\begin{longtable}[]{@{}rrr@{}}
\caption{Active-live KV accumulation under chunked
prefill.}\tabularnewline
\toprule\noalign{}
Chunk & Newly scheduled blocks & Active live blocks after chunk \\
\midrule\noalign{}
\endfirsthead
\toprule\noalign{}
Chunk & Newly scheduled blocks & Active live blocks after chunk \\
\midrule\noalign{}
\endhead
\bottomrule\noalign{}
\endlastfoot
1 & 20 & 20 \\
2 & 20 & 40 \\
3 & 20 & 60 \\
4 & 10 & 70 \\
\end{longtable}

With 60 protected resident blocks and 80 usable blocks, the maximum
active live footprint still violates the boundary.

\subsubsection{5.3 Q3: Does Infeasibility Force Explicit
Arbitration?}\label{q3-does-infeasibility-force-explicit-arbitration}

The footprint-density live failure exposed the active/resident boundary
directly. The intended resident compact spans required 60 blocks. A
bulky active prefill required 70 blocks. The usable pool was 80 blocks.
Therefore, preserving both resident prefixes while serving the bulky
active request required 130 blocks of state in an 80-block pool.

In the active-prefill trace, all policies produced zero resident
thresholded value in that setup. A value-density policy protected small
prefixes in its logical action log, but the subsequent 70-block active
prefill allocated through the protected cached blocks before explicit
filler pressure was even needed.

The direct BlockPool probe confirms the mechanism:

\begin{longtable}[]{@{}
  >{\raggedright\arraybackslash}p{(\linewidth - 2\tabcolsep) * \real{0.5000}}
  >{\raggedright\arraybackslash}p{(\linewidth - 2\tabcolsep) * \real{0.5000}}@{}}
\caption{BlockPool probe for the 60/70/80 feasibility
boundary.}\tabularnewline
\toprule\noalign{}
\begin{minipage}[b]{\linewidth}\raggedright
Probe
\end{minipage} & \begin{minipage}[b]{\linewidth}\raggedright
Result
\end{minipage} \\
\midrule\noalign{}
\endfirsthead
\toprule\noalign{}
\begin{minipage}[b]{\linewidth}\raggedright
Probe
\end{minipage} & \begin{minipage}[b]{\linewidth}\raggedright
Result
\end{minipage} \\
\midrule\noalign{}
\endhead
\bottomrule\noalign{}
\endlastfoot
Resident blocks cached & 60 \\
Active blocks requested & 70 \\
Usable pool & 80 \\
Active allocation from ordinary pool & 70 allocated \\
Resident blocks evicted & 50 \\
Resident blocks remaining cached & 10 \\
Protected via \texttt{BlockPool.touch} & Active allocation fails instead
of evicting residents \\
\end{longtable}

The probe shows both sides of the boundary. Without hard protection,
resident blocks are free-queue victims. With hard protection, the active
request cannot allocate 70 blocks from only 20 remaining free blocks.
Protection changes the failure mode; it does not make the impossible
fit.

The vLLM prototype implements the smallest contract subset needed to
change the failure mode:

\begin{itemize}
\tightlist
\item
  resident claim metadata attached to materialized prefix-cache blocks;
\item
  write no-admit for selected active requests;
\item
  hard protected-resident exclusion from ordinary free-block victims;
\item
  claim lifecycle events for accepted, materialized, demoted, expired,
  and harmed claims;
\item
  scheduler-visible active refusal when protected residents consume the
  required headroom.
\end{itemize}

The allocator-level capacity sweep uses the canonical 60 resident / 70
active case. The capacity-sweep heatmap shows the full sweep rather than
representative rows.

\begin{figure}
\centering
\includegraphics[width=0.95\linewidth,height=\textheight,keepaspectratio,alt={Capacity sweep for a 60-block resident claim and 70-block active request. Native and write-no-admit policies serve active work below the 130-block feasibility boundary by losing resident reusable KV. Hard resident exclusion preserves the accepted resident claim and converts infeasible active/resident coexistence into scheduler-visible refusal. At and above 130 usable blocks, active and resident KV can coexist. Source artifact: kv-residency-vllm-arbiter/artifacts/capacity\_sweep/capacity\_sweep\_results.json; reproduced with make capacity-sweep.}]{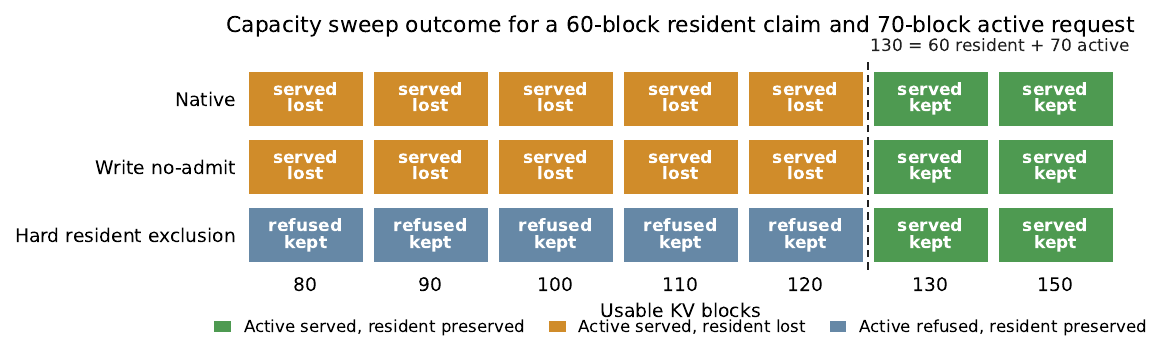}
\caption{Capacity sweep for a 60-block resident claim and 70-block
active request. Native and write-no-admit policies serve active work
below the 130-block feasibility boundary by losing resident reusable KV.
Hard resident exclusion preserves the accepted resident claim and
converts infeasible active/resident coexistence into scheduler-visible
refusal. At and above 130 usable blocks, active and resident KV can
coexist. Source artifact:
\protect\texttt{kv-residency-vllm-arbiter/artifacts/capacity\_sweep/capacity\_sweep\_results.json};
reproduced with \protect\texttt{make\ capacity-sweep}.}
\end{figure}

This sweep is the central prototype result. It does not show that the
arbiter improves production throughput or latency. It shows that the
contract is operational: below the feasibility boundary, hard resident
exclusion changes the outcome from unreported resident loss to explicit
active-side action; at the boundary, active service and resident
preservation coexist.

\subsubsection{5.4 Q4: Can An Observer Reconstruct The
Outcome?}\label{q4-can-an-observer-reconstruct-the-outcome}

Yes, for the prototype traces. The artifact includes a conformance suite
in addition to the prototype patch. The suite generates fresh direct
vLLM traces for the canonical allocator cases and combines them with
executable materialization tests from the MicroRuntime.

The cited artifact reports seven executable trace/materialization checks
plus one executable capability-classification check. The first seven
checks exercise MicroRuntime or direct vLLM traces; the final check is
not a runtime probe, but a soundness classification over backend
capabilities.

The checks are:

\begin{itemize}
\item
  \textbf{L1: No accepted claim, no claim harm.} Ordinary cached-prefix
  eviction may remove resident blocks, but cannot be labeled claim harm
  without claim acceptance. The native 60/70/80 trace has 50 resident
  victims, zero accepted claims, and zero claim-harm events.
\item
  \textbf{L2: Write no-admit separation.} Future reusable admission
  denial is separate from active-live allocation. The trace serves
  active work, denies future reuse, and still leaves 50 resident
  victims.
\item
  \textbf{L3: Hard-claim infeasibility.} Accepted hard claims under
  infeasible pressure require explicit active-side action or
  release/harm telemetry. The trace includes claim acceptance,
  materialization, and active refusal attributed to
  \texttt{claim:resident}, with 130 resident-plus-active blocks, 80
  usable blocks, and a 50-block capacity shortfall.
\item
  \textbf{L4/L5: Demotion or expiry before loss.} Later block loss after
  demotion or expiry is not claim harm. The demotion and expiry traces
  each record the lifecycle transition, zero claim-harm events, and 50
  block-loss-after-release events.
\item
  \textbf{L6: Materialization failure.} Surviving blocks can still fail
  the useful-prefix predicate. The synthetic predicate row has 59
  surviving blocks, zero leading blocks, a 60-block requirement, and
  failed materialization.
\item
  \textbf{L7: Trace reconstruction.} An external observer can
  reconstruct acceptance, materialization, active conflict, blocking
  claims, and final outcome. The L3 trace contains accepted and
  materialized lifecycle events plus attributed active refusal.
\item
  \textbf{C1: Backend capability classification.} Soft priority must not
  be treated as a sound hard-claim lowering. The capability check marks
  hard-protected claims lowered to soft priority as unsound and assigns
  soft priority to an approximate conformance class.
\end{itemize}

The suite provides direct causal attribution for the active refusal
event. The canonical refusal event names the blocking resident claim and
records the capacity proof:

\begin{Shaded}
\begin{Highlighting}[]
\FunctionTok{\{}
  \DataTypeTok{"event"}\FunctionTok{:} \StringTok{"active\_request\_refused"}\FunctionTok{,}
  \DataTypeTok{"request\_id"}\FunctionTok{:} \StringTok{"active"}\FunctionTok{,}
  \DataTypeTok{"blocking\_claim\_ids"}\FunctionTok{:} \OtherTok{[}\StringTok{"claim:resident"}\OtherTok{]}\FunctionTok{,}
  \DataTypeTok{"protected\_resident\_blocks"}\FunctionTok{:} \DecValTok{60}\FunctionTok{,}
  \DataTypeTok{"active\_live\_blocks\_required"}\FunctionTok{:} \DecValTok{70}\FunctionTok{,}
  \DataTypeTok{"resident\_plus\_active\_blocks"}\FunctionTok{:} \DecValTok{130}\FunctionTok{,}
  \DataTypeTok{"usable\_blocks"}\FunctionTok{:} \DecValTok{80}\FunctionTok{,}
  \DataTypeTok{"capacity\_shortfall\_blocks"}\FunctionTok{:} \DecValTok{50}\FunctionTok{,}
  \DataTypeTok{"feasibility"}\FunctionTok{:} \StringTok{"infeasible\_preserve\_resident\_and\_active"}
\FunctionTok{\}}
\end{Highlighting}
\end{Shaded}

The scheduler-path pressure run exercises \texttt{vllm.LLM.generate},
not only direct \texttt{BlockPool} calls. It uses a smaller SmolLM2
setup with constrained KV memory, a protected resident claim, and a
distinct active request. The active request is deferred by the scheduler
gate and then returned as a controlled terminal output:

\begin{longtable}[]{@{}ll@{}}
\caption{Live scheduler-path pressure observations.}\tabularnewline
\toprule\noalign{}
Field & Observation \\
\midrule\noalign{}
\endfirsthead
\toprule\noalign{}
Field & Observation \\
\midrule\noalign{}
\endhead
\bottomrule\noalign{}
\endlastfoot
Runtime path & \texttt{vllm.LLM.generate} \\
Model & \texttt{SmolLM2-135M-Instruct} \\
Protected resident materialized events & 40 \\
Active deferred events & 1 \\
Active refused events & 1 \\
Active stop reason & \texttt{protected\_resident\_capacity\_refused} \\
Resident claim accepted events & 1 \\
Blocking claim ids & \texttt{{[}"claim:live-resident"{]}} \\
Capacity proof & \texttt{40\ +\ 46\ =\ 86\ \textgreater{}\ 68},
shortfall \texttt{19} blocks \\
\end{longtable}

This is an analogous live scheduler-path pressure case, not the exact
60/70/80 allocator case. Its role is to show that the explicit
active-side action can propagate through vLLM's serving path.

Prefix-cache scheduler runs with SmolLM2 and Qwen2.5-Coder-7B-Instruct
show that repeated prompts receive cached-token hits and lower TTFT
after the first request. For example, the Qwen run reports \texttt{464}
cached tokens on repeated prompts and TTFT dropping from about
\texttt{0.345s} on the first request to about \texttt{0.042s} and
\texttt{0.035s} on repeats. These TTFT rows motivate why resident prefix
survival can matter, but they are not evidence that the arbiter improves
end-to-end serving performance.

\subsubsection{5.5 Boundary: Runtime Primitives Are Fragments Of The
Contract}\label{boundary-runtime-primitives-are-fragments-of-the-contract}

The prior-art boundary table in Section 2.2 anchors the final claim.
Existing runtimes do have relevant primitives, and several are plausible
lowering targets for this contract. The open question is whether those
primitives expose the full contract as one coherent claim surface:
accepted-claim lifecycle, materialization predicate, active/resident
conflict outcome, and claim-level harm/refusal telemetry.

This is why the paper is viable under the mechanisms framing. The claim
is not that modern runtimes lack retention primitives. The claim is that
retention primitives need an active/resident ownership contract to be
semantically complete. With that boundary fixed, existing primitives can
be evaluated by whether they satisfy the contract natively,
approximately, or only with adapter support.

\subsection{6. Anomalies Resolved}\label{anomalies-resolved}

The active/resident framing explains several observations that a pure
``better cache eviction'' framing does not explain cleanly.

\textbf{No-admit stops future reuse but residents still die.} This is
not a contradiction. No-admit controls future reusable admission.
Resident death happens during active live allocation from the same
physical pool.

\textbf{Naive fair share retained many blocks but produced zero useful
value.} This is not because fair share retained no KV. It retained the
wrong shape of KV: fragments below the leading-prefix thresholds that
carried policy-defined value.

\textbf{Hard resident protection can make active allocation fail.} That
is the expected result when
\texttt{resident\ +\ active\ \textgreater{}\ usable}. Protection
converts unreported resident loss into an explicit active-side refusal,
deferral, or need for another action.

\textbf{Chunking did not solve the full-attention case.} Chunking a
prefill schedule is not equivalent to bounding live KV. Under full
attention, previous chunks remain live unless the runtime introduces
offload, recomputation, or another memory-reduction mechanism.

\textbf{Strong prior art defines the lowering targets.} TensorRT-LLM,
SGLang/HiCache, Dynamo, Continuum, KVFlow, Pie, Marconi, and
Mooncake-style storage are exactly the kind of mechanisms a real system
needs. The contract defines the obligations those mechanisms would need
to satisfy around accepted resident claims, useful materialization
predicates, feasibility checks, action choice, and claim-level
telemetry.

\subsection{7. Prototype And Proposed Runtime
Surface}\label{prototype-and-proposed-runtime-surface}

A runtime-level active/resident contract should expose at least the
following concepts.

\begin{longtable}[]{@{}
  >{\raggedright\arraybackslash}p{(\linewidth - 2\tabcolsep) * \real{0.5000}}
  >{\raggedright\arraybackslash}p{(\linewidth - 2\tabcolsep) * \real{0.5000}}@{}}
\caption{Proposed runtime surface for active/resident KV
claims.}\tabularnewline
\toprule\noalign{}
\begin{minipage}[b]{\linewidth}\raggedright
Field or hook
\end{minipage} & \begin{minipage}[b]{\linewidth}\raggedright
Purpose
\end{minipage} \\
\midrule\noalign{}
\endfirsthead
\toprule\noalign{}
\begin{minipage}[b]{\linewidth}\raggedright
Field or hook
\end{minipage} & \begin{minipage}[b]{\linewidth}\raggedright
Purpose
\end{minipage} \\
\midrule\noalign{}
\endhead
\bottomrule\noalign{}
\endlastfoot
Cache-equivalence identity & Defines when the claimed prefix object is
actually reusable: model, tokenizer/hash domain, namespace, block size,
adapter, and KV format. \\
Resident protected footprint & How many blocks or bytes are being
claimed for future reuse. \\
Useful-prefix materialization rule & What shape of survival carries
value, for example a leading prefix threshold. \\
Active live footprint estimate & How much KV the active request must
hold while being served. \\
Usable KV capacity and headroom & The physical boundary for resident
plus active KV. \\
Protection mode & Whether the claim is soft priority, hard protected,
demotable, offloadable, expiring, or best effort. \\
Future reusable admission decision & Whether newly produced active KV
should be admitted for later reuse. \\
Claim-level telemetry & Whether a claim was accepted, refused, demoted,
evicted, offloaded, or harmed, with value loss. \\
Optional policy inputs & Value, confidence, budgets, deadlines, and
route/offload costs can guide admission controllers, but they are not
required fields in the conformance schema. \\
\end{longtable}

The vLLM prototype implements a minimal subset of this surface: hard
resident victim exclusion, write no-admit as a separate admission
control, claim lifecycle events, and scheduler-visible active refusal.
It does not implement a stable public API, offload, routing, recompute,
learned prediction, or an optimized policy.

That small surface already distinguishes three cases common vocabulary
tends to blur:

\begin{enumerate}
\def\labelenumi{\arabic{enumi}.}
\tightlist
\item
  Active request served and resident claim preserved.
\item
  Active request served by harming, relaxing, or offloading resident
  claim.
\item
  Active request refused or delayed because resident claim is protected.
\end{enumerate}

\subsection{8. Falsifiable Predictions}\label{falsifiable-predictions}

The mechanism makes falsifiable predictions, each tested by the cited
artifacts.

\begin{longtable}[]{@{}
  >{\raggedright\arraybackslash}p{(\linewidth - 2\tabcolsep) * \real{0.5000}}
  >{\raggedright\arraybackslash}p{(\linewidth - 2\tabcolsep) * \real{0.5000}}@{}}
\caption{Falsifiable predictions and observed outcomes.}\tabularnewline
\toprule\noalign{}
\begin{minipage}[b]{\linewidth}\raggedright
Prediction
\end{minipage} & \begin{minipage}[b]{\linewidth}\raggedright
Outcome
\end{minipage} \\
\midrule\noalign{}
\endfirsthead
\toprule\noalign{}
\begin{minipage}[b]{\linewidth}\raggedright
Prediction
\end{minipage} & \begin{minipage}[b]{\linewidth}\raggedright
Outcome
\end{minipage} \\
\midrule\noalign{}
\endhead
\bottomrule\noalign{}
\endlastfoot
Naive fair share can retain blocks while failing thresholded
leading-prefix value. & Confirmed in the controlled carrier trace:
\texttt{480\ /\ 320\ /\ 304} cached tokens but value \texttt{0}. \\
Complete-prefix or value-density policies can outperform raw
block-sharing under thresholded value. & Confirmed:
\texttt{640\ /\ 640\ /\ 0} and value \texttt{18}. \\
Write no-admit prevents bulky future reuse. & Confirmed: bulky repeat
fell from \texttt{1120} cached tokens to \texttt{0}. \\
Write no-admit alone does not protect residents from active allocation
pressure. & Confirmed: resident \texttt{small\_hot} and
\texttt{small\_warm} both returned \texttt{0}. \\
If protected resident and active live KV exceed usable capacity, another
action is required. & Confirmed by live failure, direct BlockPool probe,
and MicroRuntime arbiter. \\
Ordinary chunk scheduling does not bound live KV under full attention. &
Confirmed by MicroRuntime active-live accumulation analysis. \\
Hard protected-resident exclusion should convert sub-boundary pressure
into active-side refusal. & Confirmed in the allocator-level capacity
sweep below 130 usable blocks. \\
The explicit active-side action should be observable in a real scheduler
path. & Confirmed in a SmolLM2 \texttt{vllm.LLM.generate} pressure run
with \texttt{protected\_resident\_capacity\_refused}. \\
\end{longtable}

\subsection{9. Claim Boundaries And
Limitations}\label{claim-boundaries-and-limitations}

The evidence supports the following claims.

\begin{longtable}[]{@{}
  >{\raggedright\arraybackslash}p{(\linewidth - 2\tabcolsep) * \real{0.5000}}
  >{\raggedright\arraybackslash}p{(\linewidth - 2\tabcolsep) * \real{0.5000}}@{}}
\caption{Claims supported by the current evidence.}\tabularnewline
\toprule\noalign{}
\begin{minipage}[b]{\linewidth}\raggedright
Supported claim
\end{minipage} & \begin{minipage}[b]{\linewidth}\raggedright
Basis
\end{minipage} \\
\midrule\noalign{}
\endfirsthead
\toprule\noalign{}
\begin{minipage}[b]{\linewidth}\raggedright
Supported claim
\end{minipage} & \begin{minipage}[b]{\linewidth}\raggedright
Basis
\end{minipage} \\
\midrule\noalign{}
\endhead
\bottomrule\noalign{}
\endlastfoot
Future-reuse hints cannot be treated as unconditional commands. & The
active/resident capacity inequality can be physically infeasible. \\
Retained KV blocks are not the right value unit. & Controlled carrier
trace shows retained fragments with zero thresholded value. \\
Useful prefix reuse depends on materialization shape. & Ledger-checked
leading-prefix survival distinguishes naive and complete policies. \\
Active live KV can evict resident reusable KV before write admission
matters. & Public no-admit trace, native BlockPool trace, and vLLM
prototype patch. \\
No-admit alone is insufficient for resident protection. & Bulky repeat
reuse is eliminated while residents still die. \\
Chunking compute is not the same as bounding active live KV. &
MicroRuntime full-attention accumulation. \\
A minimal active/resident contract can be implemented inside vLLM. &
Prototype capacity sweep and live scheduler pressure run. \\
Prefix reuse has serving-visible TTFT effects. & Repeated-prefix
scheduler runs show cached-token hits and lower TTFT after the first
request. \\
Existing runtimes and orchestration systems expose fragments, not a
unified active/resident claim contract. & vLLM, SGLang, TensorRT-LLM,
Dynamo, Continuum, KVFlow, Pie, Marconi, and Mooncake comparison. \\
\end{longtable}

The evidence does not support the following claims.

\begin{longtable}[]{@{}
  >{\raggedright\arraybackslash}p{(\linewidth - 2\tabcolsep) * \real{0.5000}}
  >{\raggedright\arraybackslash}p{(\linewidth - 2\tabcolsep) * \real{0.5000}}@{}}
\caption{Claims not supported by the current evidence.}\tabularnewline
\toprule\noalign{}
\begin{minipage}[b]{\linewidth}\raggedright
Unsupported claim
\end{minipage} & \begin{minipage}[b]{\linewidth}\raggedright
Why it is out of scope
\end{minipage} \\
\midrule\noalign{}
\endfirsthead
\toprule\noalign{}
\begin{minipage}[b]{\linewidth}\raggedright
Unsupported claim
\end{minipage} & \begin{minipage}[b]{\linewidth}\raggedright
Why it is out of scope
\end{minipage} \\
\midrule\noalign{}
\endhead
\bottomrule\noalign{}
\endlastfoot
Production LLM serving speedup & No production traffic benchmark or
latency study is presented. \\
p95 or p99 latency improvement & The measurements are mechanism
counterexamples, not latency evaluations. \\
Universal allocator superiority & The proposed contract names actions;
it does not prove one global policy. \\
Runtime superiority over TensorRT-LLM, SGLang, or vLLM & The paper
compares abstractions and mechanisms, not optimized systems. \\
Upstream-ready API & The prototype is env-var driven and patch-level; it
is not a stabilized API design. \\
Learned prediction validation & Reuse value and confidence are inputs,
not learned predictors evaluated here. \\
Empirical cross-runtime generality & vLLM is the live runtime; SGLang
and TensorRT-LLM are source/design comparators. \\
Multi-tenant fairness & Fairness across tenants is orthogonal and not
solved here. \\
\end{longtable}

\subsection{10. Discussion}\label{discussion}

This paper should be read as a systems-mechanisms paper rather than a
production-performance evaluation. A performance study would need
workload distributions, latency and throughput measurements,
memory-overhead accounting, policy tuning, cross-model replication, and
comparisons against optimized runtime knobs. The contribution here is
narrower: it identifies an abstraction boundary that current terminology
blurs, demonstrates that the boundary appears in a real runtime, shows
why simpler primitives fail, and implements a minimal contract subset
that changes the conflict outcome.

The practical value of the contract is diagnostic as much as
prescriptive. Once resident reusable KV, active live KV, and future
reusable admission are separate concepts, several engineering decisions
become explicit:

\begin{itemize}
\tightlist
\item
  A write no-admit hook should not be sold as resident protection.
\item
  A resident priority hint should specify whether it is soft ranking,
  hard victim exclusion, or a claim that can refuse active work.
\item
  A chunked prefill scheduler should report whether it bounds live KV or
  only schedules compute in smaller bursts.
\item
  Offload should say whether it demotes residents, active state, or
  future reusable admissions.
\item
  Telemetry should attribute harm to resident claims, not just emit
  block store and remove events.
\end{itemize}

The TensorRT-LLM and Dynamo comparisons illustrate this boundary. A
strong runtime or orchestration layer can contain most of the
ingredients and still motivate the contract, because the paper's unit is
not a single primitive. The unit is the semantic path from prediction to
accepted claim, protected materialization, active conflict resolution,
and telemetry.

The TTFT evidence should be read narrowly. It shows that prefix reuse is
visible at the serving interface, so preserving resident prefixes can
matter. It does not show that the arbiter improves overall inference
performance, because the arbiter can intentionally defer or refuse
active work.

\subsection{11. Conclusion}\label{conclusion}

Predictive KV residency is a claim on future useful materialization, not
merely a suggestion to rank cached blocks. When active live KV and
resident reusable KV do not fit in the same usable pool, a runtime must
arbitrate ownership. Retention priority, write no-admit, chunked
prefill, cache replacement, routing, storage, and offload are all useful
pieces, but each is incomplete when treated as the whole mechanism.

Controlled vLLM traces demonstrate the counterexample in the controlled
prototype setting. The MicroRuntime isolates the contract semantics. The
vLLM prototype shows the minimal contract behavior in allocator-level
and scheduler-path settings. The runtime and prior-art audits bound the
novelty claim. Together, these artifacts support the central claim:
existing retention primitives should be understood as fragments of a
resident-claim contract for active/resident KV arbitration.

\subsection{References}\label{references}

{[}1{]} Woosuk Kwon et al., ``Efficient Memory Management for Large
Language Model Serving with PagedAttention,'' arXiv,
\url{https://arxiv.org/abs/2309.06180}.

{[}2{]} vLLM prefix caching documentation,
\url{https://docs.vllm.ai/en/v0.17.0/design/prefix_caching/}.

{[}3{]} vLLM issue ``{[}RFC{]}: Context-Aware KV-Cache Retention API
(Prioritized Evictions),''
\url{https://github.com/vllm-project/vllm/issues/37003}.

{[}4{]} NVIDIA Developer Blog, ``Introducing New KV Cache Reuse
Optimizations in NVIDIA TensorRT-LLM,''
\url{https://developer.nvidia.com/blog/introducing-new-kv-cache-reuse-optimizations-in-nvidia-tensorrt-llm/}.

{[}5{]} TensorRT-LLM KV cache documentation,
\url{https://nvidia.github.io/TensorRT-LLM/features/kvcache.html}.

{[}6{]} TensorRT-LLM useful runtime flags documentation,
\url{https://nvidia.github.io/TensorRT-LLM/performance/performance-tuning-guide/useful-runtime-flags.html}.

{[}7{]} SGLang HiCache design documentation,
\url{https://docs.sglang.io/docs/advanced_features/hicache_design}.

{[}8{]} SGLang server arguments documentation,
\url{https://docs.sglang.io/docs/advanced_features/server_arguments}.

{[}9{]} NVIDIA Dynamo agentic workflow documentation,
\url{https://docs.nvidia.com/dynamo/dev/user-guides/agents}.

{[}10{]} NVIDIA Dynamo SGLang agentic workload documentation,
\url{https://docs.nvidia.com/dynamo/dev/backends/sg-lang/agentic-workloads}.

{[}11{]} H. Li et al., ``Continuum: Efficient and Robust Multi-Turn LLM
Agent Scheduling with KV Cache Time-to-Live,'' arXiv,
\url{https://arxiv.org/abs/2511.02230}.

{[}12{]} Z. Pan et al., ``KVFlow: Efficient Prefix Caching for
Accelerating LLM-Based Multi-Agent Workflows,'' arXiv,
\url{https://arxiv.org/abs/2507.07400}.

{[}13{]} In Gim et al., ``Pie: A Programmable Serving System for
Emerging LLM Applications,'' SOSP 2025 / arXiv,
\url{https://arxiv.org/abs/2510.24051}.

{[}14{]} Rui Pan et al., ``Marconi: Prefix Caching for the Era of Hybrid
LLMs,'' Proceedings of Machine Learning and Systems 7 (MLSys 2025),
\url{https://proceedings.mlsys.org/paper_files/paper/2025/hash/7c180af017258d239bac6248d1eb26ac-Abstract-Conference.html}.

{[}15{]} vLLM project blog, ``vLLM x Mooncake: KV Cache-Centric
Disaggregated Architecture for LLM Serving,''
\url{https://vllm.ai/blog/2026-05-06-mooncake-store}.

\subsection{Appendix A: Reproducibility
Inventory}\label{appendix-a-reproducibility-inventory}

This inventory records the public, commit-pinned artifacts used by the
manuscript. Claims in the paper are tied to the public repositories,
generated artifact files, and commands listed below; no result depends
on private run capsules.

The live runtime evidence uses controlled vLLM 0.19.1 runs with the
prototype patch below, plus a companion MicroRuntime model. The most
deterministic checks are the MicroRuntime pytest suite, direct BlockPool
probes, capacity sweep, and conformance reconstruction. TTFT traces are
serving-visible motivation only and can vary with hardware, model
loading, and scheduler configuration.

Primary public repositories:

\begin{itemize}
\tightlist
\item
  MicroRuntime model:
  \href{https://github.com/gustavgauge/kv-residency-microruntime/tree/0dca4044ec8d813e3843f0788062c3ae463678d6}{\texttt{gustavgauge/kv-residency-microruntime@0dca404}}.
\item
  vLLM arbiter artifact:
  \href{https://github.com/gustavgauge/kv-residency-vllm-arbiter/tree/816541dd220d462015a2f945d2562b92310e2bd4}{\texttt{gustavgauge/kv-residency-vllm-arbiter@816541d}}.
\end{itemize}

Artifact-to-result map:

\begin{itemize}
\item
  \textbf{Q1 materialization predicate.} Artifacts:
  \href{https://github.com/gustavgauge/kv-residency-microruntime/blob/0dca4044ec8d813e3843f0788062c3ae463678d6/tests/test_materialization_fidelity.py}{materialization
  fidelity test} and
  \href{https://github.com/gustavgauge/kv-residency-microruntime/blob/0dca4044ec8d813e3843f0788062c3ae463678d6/scripts/materialization_report.py}{materialization
  report}. Commands: run the materialization fidelity pytest and the
  materialization report script.
\item
  \textbf{Q2 write no-admit boundary.} Artifacts:
  \href{https://github.com/gustavgauge/kv-residency-vllm-arbiter/blob/816541dd220d462015a2f945d2562b92310e2bd4/artifacts/no_admit/no_admit_60_70_80_summary.json}{no-admit
  summary},
  \href{https://github.com/gustavgauge/kv-residency-vllm-arbiter/blob/816541dd220d462015a2f945d2562b92310e2bd4/artifacts/conformance/L2_write_no_admit_summary.json}{L2
  conformance summary}, and
  \href{https://github.com/gustavgauge/kv-residency-microruntime/blob/0dca4044ec8d813e3843f0788062c3ae463678d6/tests/test_active_prefill_admission.py}{active-prefill
  admission test}. Commands: \texttt{make\ no-admit-probe},
  \texttt{make\ conformance}, and the active-prefill pytest.
\item
  \textbf{Q3 60/70/80 native boundary.} Artifacts:
  \href{https://github.com/gustavgauge/kv-residency-vllm-arbiter/blob/816541dd220d462015a2f945d2562b92310e2bd4/artifacts/native_blockpool/native_blockpool_60_70_80_summary.json}{native
  BlockPool summary} and
  \href{https://github.com/gustavgauge/kv-residency-vllm-arbiter/blob/816541dd220d462015a2f945d2562b92310e2bd4/artifacts/hard_claim/hard_claim_60_70_80_summary.json}{hard-claim
  summary}. Commands:
  \texttt{make\ native-blockpool-probe\ native-summary} and
  \texttt{make\ hard-claim-probe\ classify-hard-claim}.
\item
  \textbf{Q3 capacity sweep.} Artifacts:
  \href{https://github.com/gustavgauge/kv-residency-vllm-arbiter/blob/816541dd220d462015a2f945d2562b92310e2bd4/artifacts/capacity_sweep/capacity_sweep_results.json}{capacity-sweep
  results} and \href{figures/capacity-sweep-heatmap.pdf}{capacity-sweep
  heatmap}. Commands: \texttt{make\ capacity-sweep}; render the figure
  with \texttt{tools/render\_capacity\_sweep\_heatmap.py}.
\item
  \textbf{Q4 conformance and reconstruction.} Artifacts:
  \href{https://github.com/gustavgauge/kv-residency-vllm-arbiter/blob/816541dd220d462015a2f945d2562b92310e2bd4/artifacts/conformance/results.json}{conformance
  results} and
  \href{https://github.com/gustavgauge/kv-residency-vllm-arbiter/blob/816541dd220d462015a2f945d2562b92310e2bd4/artifacts/conformance/L3_hard_claim_infeasibility.jsonl}{L3
  infeasibility trace}. Command: \texttt{make\ conformance}.
\item
  \textbf{Claim demotion and expiry.} Artifacts:
  \href{https://github.com/gustavgauge/kv-residency-vllm-arbiter/blob/816541dd220d462015a2f945d2562b92310e2bd4/artifacts/claim_lifecycle/demote_summary.json}{demotion
  summary} and
  \href{https://github.com/gustavgauge/kv-residency-vllm-arbiter/blob/816541dd220d462015a2f945d2562b92310e2bd4/artifacts/claim_lifecycle/expire_summary.json}{expiry
  summary}. Commands: \texttt{make\ claim-lifecycle} and
  \texttt{make\ conformance}.
\item
  \textbf{Live scheduler pressure path.} Artifact:
  \href{https://github.com/gustavgauge/kv-residency-vllm-arbiter/blob/816541dd220d462015a2f945d2562b92310e2bd4/artifacts/live_scheduler_pressure/summary.json}{pressure
  summary}. Command: make live-scheduler-pressure.
\item
  \textbf{TTFT motivation only.} Artifacts:
  \href{https://github.com/gustavgauge/kv-residency-vllm-arbiter/blob/816541dd220d462015a2f945d2562b92310e2bd4/artifacts/live_scheduler/summary.json}{live-scheduler
  summary} and
  \href{https://github.com/gustavgauge/kv-residency-vllm-arbiter/blob/816541dd220d462015a2f945d2562b92310e2bd4/artifacts/live_scheduler/qwen25_coder_7b_summary.json}{Qwen
  summary}. Command: \texttt{make\ live-scheduler}.
\item
  \textbf{Prior-art capability boundary.} Artifact:
  \href{https://github.com/gustavgauge/kv-residency-vllm-arbiter/blob/816541dd220d462015a2f945d2562b92310e2bd4/artifacts/prior_art/prior_art_boundary.md}{prior-art
  boundary note}. Command: \texttt{make\ prior-art}.
\end{itemize}

Primary model and tests:

\begin{itemize}
\tightlist
\item
  \href{https://github.com/gustavgauge/kv-residency-microruntime/blob/0dca4044ec8d813e3843f0788062c3ae463678d6/src/kvrt/contract.py}{\texttt{src/kvrt/contract.py}}
\item
  \href{https://github.com/gustavgauge/kv-residency-microruntime/blob/0dca4044ec8d813e3843f0788062c3ae463678d6/src/kvrt/active_live.py}{\texttt{src/kvrt/active\_live.py}}
\item
  \href{https://github.com/gustavgauge/kv-residency-microruntime/blob/0dca4044ec8d813e3843f0788062c3ae463678d6/src/kvrt/arbiter.py}{\texttt{src/kvrt/arbiter.py}}
\item
  \href{https://github.com/gustavgauge/kv-residency-microruntime/blob/0dca4044ec8d813e3843f0788062c3ae463678d6/docs/hard-seeds/decision.md}{\texttt{docs/hard-seeds/decision.md}}
\item
  \href{https://github.com/gustavgauge/kv-residency-microruntime/blob/0dca4044ec8d813e3843f0788062c3ae463678d6/tests/test_resident_claim_contract.py}{\texttt{tests/test\_resident\_claim\_contract.py}}
\item
  \href{https://github.com/gustavgauge/kv-residency-microruntime/blob/0dca4044ec8d813e3843f0788062c3ae463678d6/tests/test_materialization_harness.py}{\texttt{tests/test\_materialization\_harness.py}}
\item
  \href{https://github.com/gustavgauge/kv-residency-microruntime/blob/0dca4044ec8d813e3843f0788062c3ae463678d6/tests/test_active_resident_arbiter.py}{\texttt{tests/test\_active\_resident\_arbiter.py}}
\item
  \href{https://github.com/gustavgauge/kv-residency-microruntime/blob/0dca4044ec8d813e3843f0788062c3ae463678d6/tests/test_active_prefill_admission.py}{\texttt{tests/test\_active\_prefill\_admission.py}}
\end{itemize}

Primary vLLM prototype artifacts:

\begin{itemize}
\tightlist
\item
  \href{https://github.com/gustavgauge/kv-residency-vllm-arbiter/blob/816541dd220d462015a2f945d2562b92310e2bd4/patches/vllm_resident_claim_prototype.patch}{\texttt{patches/vllm\_resident\_claim\_prototype.patch}}
\item
  \href{https://github.com/gustavgauge/kv-residency-vllm-arbiter/blob/816541dd220d462015a2f945d2562b92310e2bd4/src/kv_vllm_arbiter/conformance.py}{\texttt{src/kv\_vllm\_arbiter/conformance.py}}
\item
  \href{https://github.com/gustavgauge/kv-residency-vllm-arbiter/blob/816541dd220d462015a2f945d2562b92310e2bd4/tests/test_conformance.py}{\texttt{tests/test\_conformance.py}}
\item
  \href{https://github.com/gustavgauge/kv-residency-vllm-arbiter/blob/816541dd220d462015a2f945d2562b92310e2bd4/scripts/run_conformance_suite.py}{\texttt{scripts/run\_conformance\_suite.py}}
\item
  \href{https://github.com/gustavgauge/kv-residency-vllm-arbiter/blob/816541dd220d462015a2f945d2562b92310e2bd4/artifacts/conformance/results.json}{\texttt{artifacts/conformance/results.json}}
\item
  \href{https://github.com/gustavgauge/kv-residency-vllm-arbiter/blob/816541dd220d462015a2f945d2562b92310e2bd4/artifacts/conformance/summary.md}{\texttt{artifacts/conformance/summary.md}}
\item
  \href{https://github.com/gustavgauge/kv-residency-vllm-arbiter/blob/816541dd220d462015a2f945d2562b92310e2bd4/artifacts/conformance/L3_hard_claim_infeasibility.jsonl}{\texttt{artifacts/conformance/L3\_hard\_claim\_infeasibility.jsonl}}
\item
  \href{https://github.com/gustavgauge/kv-residency-vllm-arbiter/blob/816541dd220d462015a2f945d2562b92310e2bd4/artifacts/conformance/L3_hard_claim_infeasibility_summary.json}{\texttt{artifacts/conformance/L3\_hard\_claim\_infeasibility\_summary.json}}
\item
  \href{https://github.com/gustavgauge/kv-residency-vllm-arbiter/blob/816541dd220d462015a2f945d2562b92310e2bd4/artifacts/capacity_sweep/capacity_sweep_results.json}{\texttt{artifacts/capacity\_sweep/capacity\_sweep\_results.json}}
\item
  \href{https://github.com/gustavgauge/kv-residency-vllm-arbiter/blob/816541dd220d462015a2f945d2562b92310e2bd4/artifacts/live_scheduler_pressure/summary.json}{\texttt{artifacts/live\_scheduler\_pressure/summary.json}}
\item
  \href{https://github.com/gustavgauge/kv-residency-vllm-arbiter/blob/816541dd220d462015a2f945d2562b92310e2bd4/artifacts/claim_lifecycle/demote_summary.json}{\texttt{artifacts/claim\_lifecycle/demote\_summary.json}}
\item
  \href{https://github.com/gustavgauge/kv-residency-vllm-arbiter/blob/816541dd220d462015a2f945d2562b92310e2bd4/artifacts/claim_lifecycle/expire_summary.json}{\texttt{artifacts/claim\_lifecycle/expire\_summary.json}}
\end{itemize}

\end{document}